# $C_{58}$ on Au(111): a scanning tunneling microscopy study

Noelia Bajales[1,3,4], Stefan Schmaus[1], Toshio Miyamashi[1], Wulf Wulfhekel[1,2],

Jan Wilhelm[2], Michael Walz[2], Melanie Stendel[2], Alexej Bagrets[2], Ferdinand Evers[2],

Seyithan Ulas[3], Bastian Kern[3], Artur Böttcher*[3] and Manfred M. Kappes*[2,3]

[1]*Institute of Physics, Karlsruhe Institute of Technology (KIT), Wolfgang-Gaede-Str. 1, D-76131 Karlsruhe, Germany;*

[2]*Institute of Nanotechnology, Karlsruhe Institute of Technology (KIT), Hermann-von-Helmholtz-Platz 1, D -76021 Eggenstein-Leopoldshafen, Germany;*

[3]*Institute of Physical Chemistry, Karlsruhe Institute of Technology (KIT), Fritz-Haber-Weg 2, 76131 Karlsruhe, Germany;*

[4] *Grupo Ciencia de Materiales, Instituto Enrique Gaviola (IFEG), 5000 Córdoba, Argentina*

**Abstract**

$C_{58}$ fullerenes were adsorbed onto room temperature Au(111) surface by low-energy (~6 eV) cluster ion beam deposition under ultrahigh vacuum conditions. The topographic and electronic properties of the deposits were monitored by means of scanning tunnelling microscopy (STM at 4.2 K). Topographic images reveal that at low coverages fullerene cages are pinned by point dislocation defects on the herringbone reconstructed gold terraces (as well as by step edges). At intermediate coverages, pinned monomers, act as nucleation centres for the formation of oligomeric $C_{58}$ chains and 2D islands. At the largest coverages studied, the surface becomes covered by 3D interlinked $C_{58}$ cages. STM topographic images of pinned single adsorbates are essentially featureless. The corresponding local densities of states are consistent with strong cage-substrate interactions. Topographic images of $[C_{58}]_n$ oligomers show a stripe-like intensity pattern oriented perpendicular to the axis connecting the cage centers. This striped pattern becomes even more pronounced in maps of the local density of states. As supported by density functional theory, DFT calculations, and also by analogous STM images previously obtained for $C_{60}$ polymers [M. Nakaya et al., J. Nanosci. Nanotechnol. 11, 2829 (2011)], we conclude that these striped orbital patterns are a fingerprint of covalent intercage bonds. For thick $C_{58}$ films we have derived a band gap of 1.2 eV from scanning tunnelling spectroscopy data, STS, confirming that the outermost $C_{58}$ layer behaves as a wide band semiconductor.

Corresponding authors: artur.boettcher@kit.edu; manfred.kappes@kit.edu

## 1. Introduction

Arguably, the birth of carbon nanoscience coincides with the discovery of the Buckminsterfullerenes: all-carbon, hollow molecular cages consisting of hexagonal and pentagonal faces /[1]/. Early mass spectra of the volatile products created by pulsed laser ablation of graphite in a helium atmosphere showed not only the well-known $C_{60}$ and $C_{70}$ cages but also smaller $C_n$ species (n-even) in still significant abundances. These smaller cages have largely been ignored over the years. Primarily this is a consequence of more difficult processability: $C_{60}(I_h)$ as well as other higher fullerenes can be efficiently solvent extracted from polydisperse fullerene soots - as generated by the standard Krätschmer-Huffman arc discharge production method /[2]/. In contrast, smaller $C_n$ species also present in these soots cannot be dissolved and as a result they have been mostly studied theoretically. For example, density functional theory (DFT) calculations have predicted their molecular structures /[3,4]/. Such small $C_n$ cages (n<60) always have "non-IPR fullerene" structures – where IPR refers to the *Isolated Pentagon Rule* initially formulated to systematize the structures of solvent extractable "conventional" cages. In the latter (e.g. $C_{60}(I_h)$), the 12 pentagonal faces are *isolated* from one another by surrounding hexagons. In contrast, non-IPR cage structures can include two adjacent pentagonal rings (2AP), chains of several pentagons (e.g. 2x3AP), heptagons (Hept), and even squares (S). To first order, strain energy increases with decreasing cage size and therefore small, non-IPR fullerene cages are less stable (in terms of binding energy per atom) than their larger IPR congeners. In addition, the non-IPR motifs act as defect sites. Both effects lead to enhanced reactivity (and much poorer processability) relative to IPR cages.

Nevertheless, stable, sublimable, monodispersed cluster materials based on pure non-IPR fullerenes have recently been made. In the first such study, A. Böttcher et al./[5,6]/ generated thin films of non-IPR $C_{58}$ by ion beam deposition. The corresponding molecular building blocks were produced under ultrahigh vacuum (UHV) by means of electron impact induced ionization/fragmentation of sublimed $C_{60}(I_h)$. The resulting $C_{58}^+$ species were separated from the parent and other fragment ions using a quadrupole mass filter and then deposited onto an inert highly oriented pyrolytic graphite, (HOPG) surface at low incident kinetic energies (low energy cluster ion beam deposition = LECBD). Other monodispersed $C_n$ films, 48 < n < 68, have since been analogously prepared /[7]/. As these and subsequent measurements have shown, many of the electronic, vibrational, thermodynamic and mechanical properties of these prototypical cluster materials depend strongly on the size of the building blocks used /[8]/.

Whereas monodispersed non-IPR fullerene materials are now routinely accessible, their molecular composition still remains unclear. This has to do with the enormous variety of cage isomers which are in principle accessible at any given cluster nuclearity. For example, it is not known whether the $C_{3v}$ symmetry $C_{58}^{+}$ cage ($C_{58}(C_{3v}:0001)$) predicted by DFT to be the most stable isomer is in fact generated in the LECBD ion source. There is an energetically close lying $C_{58}(C_s:Hept)$) cage isomer for which the calculated difference in total energy relative to the most stable cage is ~10.5 kJ/mol for neutral and ~4.2 kJ/mol for the cationic cage /[9],[10]/. Consequently, *both isomers* might be present in the incident ion beam. Furthermore, interaction of the (hot) $C_n^+$ projectile with the substrate could change the final molecular structure of the cage on the surface relative to that in gas-phase. In particular, impact is associated not only with surface-mediated dissipation of the ions´ primary kinetic energy /[21]/ but also by surface-harpooning induced neutralization. Both processes release significant amounts of energy into the projectile itself.

Recent atomic force microscopy (AFM) studies of the deposits generated by LECBD of non-IPR $C_n$ onto HOPG surfaces have shown that surface temperature and primary kinetic energy can significantly influence the topography of the resulting island films on mesoscopic length scales. The dendritic islands observed have been rationalized in terms of on-surface cage-cage reactions to form stable $[C_n]_m$ oligomeric chains. However, so far it has not been possible to determine the corresponding (intramolecular) structures on an atomic level /[7]/.

In this study we have attempted to obtain such information. Specifically, we have applied low-temperature scanning tunnelling microscopy (4.2 K-STM) to image individual $C_{58}$ cages and oligomers deposited gently onto an inert Au(111) surface. The choice of Au(111) as a substrate was motivated by its well-known herringbone reconstruction/[11]/. In analogy to measurements of Ni/Au(111) /[12]/ it was expected that the herringbone surface reconstruction and in particular its characteristic point dislocation sites can act as traps to immobilize individual $C_{58}$ cages – thus facilitating STM probes without perturbations due to lateral cage-cage interactions.

The focus here has been placed on the low coverage regime for which imaging of individual cages as well as small $[C_{58}]_m$ aggregates is possible. Furthermore, the electronic structure of both occupied and unoccupied cage states can be studied by scanning tunnelling spectroscopy, STS. In order to qualitatively interpret the measurements, we have also simulated STM topographic images by means of DFT calculations using various structural models: individual $C_{58}$ cages and dimers in different surface configurations and orientations. Several striking features of the STM data provide clear evidence of soft-landing mediated oligomerization of carbon cages.

## 2. Techniques and Methods

### 2.1 *Low-energy cluster ion beam deposition*

Deposition of mass-selected $C_{58}^+$ cluster cations was carried out in an ultrahigh vacuum system with a base pressure better than 5 x$10^{-10}$ mbar. The $C_{58}$ cluster ions were generated by electron-impact-induced ionization, heating and fragmentation of $C_{60}$ fullerenes. After mass selection using a quadrupole mass filter, a collimated beam of $C_{58}^+$ was electrostatically directed towards an atomically clean Au(111) surface held at room temperature. Deposits and films were grown by low energy cluster ion beam deposition, LECBD /[13]/. Details have been described elsewhere /[5,6]/. In all experiments, the same mean incident kinetic energy of $C_{58}^+$ of ~ 6 eV was chosen. Previous deposition studies using HOPG substrates have demonstrated that under such low incident kinetic energy conditions, fragmentation of the impinging cages can be excluded. The films were grown at a nearly constant deposition rate of ~3 x $10^9$ cages per second per 6.7 mm$^2$ surface area. Mass spectra of the incident ion beam under the mass selection conditions used for deposition, revealed no measurable overlap between neighbouring constituents, $C_{56}^+$, $C_{58}^+$ and $C_{60}^+$, i.e. the ion beam purity of incident $C_{58}^+$ was greater than 99.99%. Similarly, more highly-resolved mass spectra showed no measurable contamination by derivatives such as $C_{58}H^+$, $C_{58}CO^+$, $C_{58}OH^+$, etc.. The number of cages deposited per surface area has been quantified by monitoring the neutralization current (picoamperemeter Keithley). A saturated monolayer, 1MLE, is reached at lateral density of $10^{14}$ $C_{58}$/cm$^{-2}$ corresponding to an actual ion dose of 1.2 x $10^{-6}$ C.

### 2.2 *STM and STS*

Atomically clean Au(111) substrates were prepared by repeated sputtering/annealing cycles /[14]/. The quality of the resulting surfaces was checked by ultraviolet photoemission spectroscopy and X-ray photoionization spectroscopy (UPS and XPS). In particular the gold surface state in the valence band (SS, $E_B$~390 meV /[15]/), a very sensitive measure of the periodic atomic alignment, was followed as a function of the number of sputtering/annealing cycles. Saturation of this signal was taken to indicate a clean substrate.

After deposition, substrates were transferred into another UHV chamber equipped with a home-built scanning tunnelling microscope operating at liquid helium temperature - by means of a UHV suitcase. The deposition and the transfer were performed at room temperature and all STM/STS measurements were subsequently conducted at a sample temperatures of 4.2 K using (still) atomically clean substrates and tunnelling tips precleaned by Ar$^+$ sputtering and annealing. The topographic images were taken in the constant current mode, i.e. an isodensity surface $z_{\rho*}(x,y)$ ($\rho*$=const) was determined at every lateral (x,y)-tip position.

Constant current topography and corresponding dI/dV values were acquired simultaneously by superimposing a weak sinusoidal modulation onto the bias potential (lock-in detection). This allows the local density of states, LDOS (~dI/dV(I/V)) to be determined at desired (x,y)-positions. The entire LDOS($E_B$-$E_F$) in the valence region around the Fermi level was acquired by performing LDOS measurements at slowly linearly increasing bias voltage $V_B$ (whereby the electron binding energy is given by $E_B=eV_B$). LDOS(x,y)-maps were recorded for specific, intentionally chosen HOMO or LUMO states by monitoring the LDOS while laterally scanning over the surface at constant bias /[16]/.

## 2.3 *DFT calculations*

For comparison with experiment, the electronic structure and STM images of $C_{58}$/Au(111) and $(C_{58})_{2,3}$/Au(111) were calculated using an approach which is described in detail elsewhere /[17]/ (see also appendix). First, the electronic structure of a set of "extended molecules" comprising fullerene cage(s) bound in various orientations to model gold cluster substrates were calculated at the DFT level with the quantum chemistry package TURBOMOLE /[18]/. Then, electron transmission probabilities into/from these "extended molecules" were calculated in the tunnelling limit for a model STM tip held at various positions above the fullerene(s) - using a non-equilibrium Green's functions (NEGF) technique as implemented in the AITRANSS module (Ab Initio TRANSport Simulations /[17,18]/).

The DFT calculations were performed using the generalized gradient approximation (GGA, BP86 functional) together with contracted Gaussian-type basis sets of split-valence (def-SVP) and triple-zeta quality (def-TZVP) including polarization functions. Corresponding Coulomb-fitting basis sets were applied within the resolution of identity (RI) approximation. In the "extended molecules", dispersion force contributions to the binding of $C_{58}$ to the model Au cluster have been explicitly taken into account by including DFT-D3-type Grimme corrections. Structural relaxations have been performed for all $C_{58}$ cages (in their various orientations). $C_{58}$ dimer geometries were relaxed with the substrate binding distance (as previously calculated for $C_{58}$ monomers) held constant. The model gold surface itself (see below) was not relaxed. In particular, no attempt was made to model pinning of $C_{58}$ on point defects in the herringbone reconstruction.

The "extended molecules" considered were based on the two lowest energy $C_{58}$ isomers: $C_{58}(C_{3v})$ and $C_{58}(C_s)$ /[3, 4, 19]/. The unreconstructed Au(111) surface has been modelled by a finite, two-layer cluster consisting of 45 gold atoms total. The topmost layer (surface) is constituted by 27 hexagonally densely packed gold atoms. Overall 18 different orientations of the $C_{58}(C_{3v})$ cage and 9 orientations of the $C_{58}(C_s)$ cage (relative to the gold substrate) were screened. These contact geometries differed by the orientation of the heptagonal, hexagonal or pentagonal faces selected to lie on top of the hexagonally packed gold atoms of the (111) surface layer. The contacting faces were in turn selected (according to their location in the Schlegel diagram of the fullerene cage) so as to be in close proximity to one or more non-IPR sites. The most stable bonding configuration was found for a $C_{58}(C_s)$ cage oriented with its heptagonal face pointing towards the surface. The $C_{58}(C_{3v})$ cage prefers an orientation in which a hexagonal face adjacent to a 2AP reaction centre lies on top of the surface.

The choice of dimer models for simulating STM images was guided by three considerations: (1) only dimers consisting of the most stable $C_{58}(C_s)$ and $C_{58}(C_{3v})$ monomers were taken into account, (2) the first of the component monomers was positioned on the surface so as to maximize its adsorption energy (see above), (3) the second cage was oriented to allow covalent interlinking of the two cages via non-IPR sites. Four plausible configurations of the most stable $C_{58}(C_s)$-$C_{58}(C_s)$ dimers as well as two $C_{58}(C_{3v})$-$C_{58}(C_{3v})$ dimer configurations have been calculated (see appendix for more calculation details and configuration of the adsorbate-substrate complex).

## 3. Results and Discussion

### 3.1 *Surface topography*

*3.1.1 Single cages on point dislocations.* STM topographic images of clean surfaces reveal regions of both ABC (fcc) and ABA (hcp) stacking as expected for well-ordered reconstructed Au(111)[20]. The stripes of the herringbone reconstruction, marking the corresponding domain boundaries, are clearly apparent. Fig. 1a shows a representative surface region after depositing a small amount of $C_{58}^{+}$ ($C_{58}$ dose ~5 x $10^{12}$ $cm^{-2}$). Nearly circular bright spots indicate individual $C_{58}$ cages. Cages deposited onto terrace regions, appear well-aligned with a nearest-neighbour distance of ~7 nm. Apparently the impinging cages become preferentially immobilized at the elbows in the herringbone reconstruction. These are in fact constituted by point dislocation sites, Au(#). Gold atoms at step edges terminating the terraces can play a similar immobilizing role (not shown here). Apparently undercoordinated gold atoms (associated with both kinds of pinning site) are responsible for the initial reactivity of the surface towards $C_{58}$. In contrast, neither the domain walls nor the dislocation line are capable of binding the mobile cages under our deposition conditions.

Lateral alignment of the cages with respect to the domain walls, D (along [-1 -1 2]), indicates significant surface mobility of the impinging cages prior to pinning/sticking. In analogy to the picture proposed for LECBD onto the HOPG basal plane [21], we conclude that the initial hyperthermal kinetic energy of $C_{58}^+$ ions (E ~6 eV) impinging onto Au(111) becomes dissipated by: (i) conversion into surface parallel motion (E→$E_{||}$) and subsequently (ii) molecular friction (plasmon and electron-hole losses accompanying lateral sliding motion). Mobile cages can first become pinned when their lateral kinetic energy drops below the binding energy at the reconstructed sites, $E_{||} < E_b$. However, pinning during the subsequent thermal diffusion stage is more probable. Thus, as illustrated by Fig. 1a the point dislocation sites, Au(#), lead to periodic decoration of terraces in the initial deposition stages.

What is the nature of the stabilizing $C_{58}$-Au(#) binding interaction? Due to relativistic effects, Au atoms can form relatively strong bonds to carbon (e.g. $E_b$(Au-C) = 3.8 eV [22,23]). However, it is unlikely that under our room temperature deposition conditions, sufficient activation energy can be provided to break open coordinatively saturated non-IPR fullerene cages (to form strong covalent -C-Au(#) bonds). Instead, extended DFT calculations performed for (unfragmented) $C_{60}(I_h)$/Au(111) indicate an interaction comprising both v.d. Waals and "covalent-like" contributions (the latter associated with surface hybridization of the $C_{60}$-LUMO) together yielding $E_b(C_{60}$-Au)=1.9 eV [24]. Our DFT calculations indicate that $C_{58}$ behaves analogously, albeit with a stronger "covalent-like" interaction contribution due to the larger cage curvature and the presence of non-IPR "defect" sites [25]. Preliminary measurements indicate that $C_{58}$ cages immobilized on Au(111) at low submonolayer coverages cannot be desorbed by heating (up to 1300 K). Based on our previous measurements of the thermal desorption of non-IPR fullerenes from HOPG, a temperature limit of 1300 K implies a lower limit for the corresponding desorption activation energy of $C_{58}$-Au(#) of at least 3.58 eV (as estimated by assuming first-order desorption kinetics and using Redhead's formula [26] E=RT[ln(νT/β)- 3.46] at ν=$10^{13}$ s$^{-1}$, β=5K/s). This number should be treated with caution however, given that our preliminary experiments also indicate significant surface decomposition (to non desorbable products) during heating to 1300 K [8]. Nevertheless, both observations support a comparatively strong $C_{58}$-Au(#) interaction.

Figure 1a shows that apparent diameter of single pinned $C_{58}$ cages depends slightly on bias voltage - ranging from 0.7 to 0.8 nm. This value is slightly larger than that calculated for isolated $C_{58}$ (~0.7 nm) because of tip convolution effects. Whereas the imaged brightness distributions associated with individual $C_{58}$ cages are typically bell-shaped (when mapped into 3D), significant lateral asymmetries may sometimes be observed. We were never able to resolve pentagonal or hexagonal rings within the bright areas. This is in contrast to $C_{60}$ cages whose STM topography clearly exhibits intramolecular structures under analogous conditions (see Fig. 1b). Line profiles taken across $C_{58}$-Au(#) complexes reveal apparent heights ranging from 0.22 to 0.26 nm (compared to ~0.22 nm for $C_{60}$). The smooth intensity distributions observed for $C_{58}$-Au(#) at 4.2 K suggest that no non-IPR site is present in the upper part of the cage as probed by STM. We speculate that all non-IPR reaction centers point towards the Au(#) site. This is plausible given the cage structure of the lowest energy $C_{58}$ isomers: $C_{58}(C_{3v})$ and $C_{58}(C_s)$. In both cases, the reactive non-IPR sites are not homogenously distributed over the cage but instead localized on only one hemisphere /[3,4]/.

***3.1.2 Single cages on terraces.*** Topographic images have also been obtained for (minority - less than 10%) $C_{58}$ species located on free terraces and therefore pinned by sites different from Au(#). Compared to $C_{58}$-Au(#), these cages (figure 1c) appear somewhat larger (~1.2 nm). Also the corresponding "bright spots" manifest asymmetries and often show internal features. Such topographies can be attributed to adsorbed $C_{58}*$ cages oriented with some of their nIPR sites pointing towards the STM tip. Fig. 1c compares five experimental images of single $C_{58}$ cages (upper row) to DFT-based predictions (iso-density surfaces) for several orientations of $C_{58}(C_s)$ and $C_{58}(C_{3v})$ on Au(111) (lower row). Image (a) reveals a hexagonally shaped contour which can be qualitatively described by a $C_{58}(C_s)$ cage adsorbed to the surface by way of pentagonal face ($C_s$-Pent3). The image calculated for hexagon-faced $C_{58}(C_{3v})$ monomer, $C_{3v}$-Hex3b, resembles the structure seen in image (b). Several other STM images (e.g. (c), (d) and (e)) were also fit quite well by the DFT simulations (for adsorption geometry see appendix). In general, however, the intracage structures seen in the "on-terrace" topographic images are not sharp enough to clearly distinguish between different isomers and their orientations.

***3.1.3 Oligomers.*** The surface topography changes considerably with increasing $C_{58}$ dose. Fig. 2a shows a representative surface area as imaged after depositing ~5 x $10^{13}$ cages/cm$^2$ (intermediate coverage regime). Now we predominantly observe planar $C_{58}$ aggregates centered at Au(#) sites. Again, the overall deposit periodicity reflects that of the herringbone-reconstructed Au(111) surface underneath. 2D islands are aligned along the [-1-12] direction as evidenced by the still recognizable underlying herringbone structure. Apparently, individual $C_{58}$-Au(#) complexes formed during the initial deposition stages can act as nucleation sites for the capture of further (surface migrating) cages. Correspondingly, the mean free path of a surface migrating/sliding cage is governed by the distance between adjacent Au(#) nucleation sites.

In the intermediate coverage regime, we also frequently find small oligomeric aggregates localized on terrace regions (presumably an immobilized cage or a cage pinned by an atomic defect acts as the corresponding nucleation center). The STM topographic images shown in Fig. 2b illustrate some of these structures, which we assign to immobilized fullerene dimers and trimers, $[C_{58}]_{n<4}$, based on their dimensions (cf. the bright areas found in STM topographic images of isolated monomers). Specifically, the topographies shown in left panel in Figure 2b correspond to differently oriented dimers, $[C_{58}]_2$, whereas the image shown in right panel reflects a linear trimer. The line profiles along the long-axes of these objects show heights (<0.3 nm) which confirm that they correspond to (linear) oligomeric $C_{58}$ chains. Unfortunately, the topographic images do not allow us to unequivocally determine cage-center positions and intercage separations. However, the images do exhibit pronounced internal structures with striking brightness distributions. Note, that this internal structure is not like that seen for isolated $C_{60}$ cages which show surface-normal symmetry axes (fig. 1b). Instead an elongated, highly anisotropic intensity distribution is typically observed with several stripe-like local intensity maxima oriented perpendicular to the axis connecting the cage centres. In some cases, dark regions separating adjacent stripes can be clearly identified as nodal planes in the molecular orbitals of the corresponding $[C_{58}]_n$ oligomers. Interestingly, quite similar STM topographic images have been previously observed in $C_{60}$ polymers generated by covalently interlinking adjacent $C_{60}$ cages – via electron irradiation [27] or by photopolymerization [28] to generate 2+2 cycloadducts. The similarity between the HOMO topography found here for $C_{58}$ dimers and trimers and that evidenced for $C_{60}$-$C_{60}$-$C_{60}$ polymeric chains [27,28,34], supports our previous inference [5,6] that LEBCD generates $[C_{58}]_n$ oligomers stabilized by covalent intercage bonds as constituted to a considerable extent by non-IPR sites on adjacent cages (e.g. 2AP-2AP, Hept-Hept...). We speculate, that the first cage in the chain pins to the surface by way of a strongly binding orientation, whereupon the other cages in the chain become oriented (relative to the first cage) according to their non-IPR site mediated intercage bonding.

The assignment of striped topographic signatures to $C_{58}$ polymers stabilized by covalent inter-cage bonds is further supported by the simulated images of $C_{58}$-$C_{58}$ dimers as calculated for several plausible (stable) surface configurations. Fig. 2c shows one example [dimer 1: $C_{58}(C_s)$-$C_{58}(C_s)$]. The predicted iso-density surface exhibits bright stripes oriented perpendicular to the dimer axis. The shape and the lateral extension of the parallel stripes depend slightly on the tip-cage distance. Changing the relative orientation (and thus the interaction) of oligomer and surface, does not significantly influence the stripes in the simulated iso-density surfaces. Importantly, parallel intensity stripes are only predicted when including covalent C-C inter-cage bonds. Such features were obtained for all covalently stabilized dimers and trimers calculated here [17]. The cage interlinking C-C bonds in fact reflect local $sp^2 \rightarrow sp^3$ rehybridisation. This in turn forces a considerable rearrangement of the delocalized π-states constituting the bonding orbitals of the dimer (in particular in the "interface region" between cages).

In a previous study of $C_{58}^+$ LECBD onto HOPG [7] covalent inter-cage bonds were inferred to form either: (i) quickly after impact (via frictional dissipation of surface parallel kinetic energy followed by collision of a sliding cage with an already existing aggregate) or (ii) after thermalization of the deposited cage followed by surface diffusional motion. LECBD-derived $C_{58}$ submonolayers on HOPG commonly exhibit 2D dendritic islands independent of primary kinetic energy (1 eV < $E_{kin}$ < 9 eV). This suggested that the effective activation energy for intercage bond formation is less than 1 eV. The formation of predominantly covalently stabilized oligomers as observed here implies that the situation is analogous for Au(111). However, the extent to which surface-parallel kinetic energy facilitates intercage bond formation on both surfaces remains to be explored in detail.

**3.1.4 Three-dimensional growth.** After depositing several monolayers of $C_{58}$, the surface becomes covered with 3D islands separated from each other by narrow substrate regions (fig. 3 (upper panel); ~2 x $10^{14}$ cages/cm$^2$). The islands are ~15 nm wide and have overall heights of ~4 monolayers. At even larger $C_{58}$ coverages, the substrate disappears entirely and a very rough $C_{58}$ film with a mean corrugation depth of ~2 ML is observed. Thus, the LECBD-mediated growth of $C_{58}$ films on Au(111) is similar to that observed on HOPG substrate. It proceeds first via 2D cage aggregation which prevails over random adsorption and consequently leads to 3D growth of pyramidal islands which at a later stage ends with wetting of the contacting layer. Note however that thin $C_{58}$ films grown on HOPG and Au(111) have quite different surface roughness. This results from different mean free paths at the low coverage stage: ~400 nm for $C_{58}$ soft-landed onto HOPG, versus ~15 nm on Au(111). Consequently, $C_{58}$ forms compact 3D islands on Au(111) (mean island density ~$1/\Lambda$), whereas large dendritic 2D islands are still observable on HOPG terraces at the same coverage /[21]/.

STM topographic images of thick films typically show grape-like-cluster structures (see Fig. 3 (middle panel). The apparent diameter distribution of individually distinguishable cages is unexpectedly wide, ranging from 0.7 up to 1.6 nm. This results from difficulties in clearly assigning the cage perimeters. Closer inspection (two panels on the right in Fig. 3) shows that the border regions between adjacent cages generally manifest stripe-like patterns as already noted for dimers in the previous section. The grape-like structures seen for thick films typically show several different kinds of stripe-like structures. This can be rationalized in terms of variations in the number, the geometry and relative orientation of inter-cage bonds. For example, the most stable cage, $C_{58}(C_{3v})$, comprises three pairs of two adjacent pentagons, 3x2AP, equidistantly distributed over one hemisphere. 3D growth of $C_{58}(C_{3v})$-based oligomers linked only by way of these reactive 2AP sites should therefore not be completely random but instead predefined by the relative spatial orientation of the reaction centers. However, such orientational ordering would also be kinetically hindered to some degree (depending on primary kinetic energy, $E_{kin}$, surface temperature, $T_s$, ionic flux, $F^+$, charge state, surface diffusion, steric effects, etc.) and as a result, thick $C_{58}$ layers are expected to also comprise cages not covalently interlinked to their maximal possible degree /[29]/.

In summary, growth proceeds by: (1) pinning at point dislocation sites, Au(#); (2) lateral growth of 2D islands as nucleated primarily by Au(#) sites (and mediated by the formation of covalent bonds involving reactive nIPR sites) and (3) formation of multilayered films via the growth of 3D grains.

**3.2** *Local densities of states*

***3.2.1 Bandgap of thick films.*** Scanning tunnelling spectroscopy (STS) was used to obtain the local densities of states in the valence-band region, LDOS /[30]/, for $C_{58}$ deposits of various thickness. Fig. 4a shows LDOS functions determined at several surface locations for nominally 3 ML-thick films comprising 3D aggregates of up to 5 cage layers in height. Pronounced variations were observed. Site (1) is dominated by contributions from the Au(111)-specific surface state at $E_B$(SS) ~ 390 meV which appears together with two broad side bands. All other sites probed show four common states in the vicinity of the Fermi energy (LUMO-I ($E_B$= 0.35 eV), LUMO-II ($E_B$= 0.5 eV), HOMO-I ($E_B$= -0.85 eV) and HOMO-II ($E_B$=-1.3 eV)) but with different relative intensities. Based also on the results of the previous section, these variations can be assigned to different bonding configurations of the topmost cages (possibly also to different $C_{58}$ isomers). Whereas, LDOS-peak energies can vary from site to site by ~0.1 eV, the difference E(HOMO-I)-E(LUMO-I), remains roughly constant at 1.2 eV. The latter is a first-order measure of the HOMO-LUMO gap thus confirming the previous inference from UPS measurements that thick $C_{58}$ films are wide band semiconductors /[31]/. Note, that the gap-width as determined by STS agrees well with the value of $\Delta$=1.2 +/-0.1 eV inferred from photoionization spectra of thick Cs-doped $C_{58}$ films /[7]/. Note further, that the "HOMO band" recorded for these films by UPS at room temperature is roughly 2.2 eV wide and exhibits three distinguishable peaks /[7]/. Its overall shape is consistent with the prominent STS features seen in Fig. 4a at $E_B$= -0.85, -1.15 and -1.3 eV. Similarly, STS measurements (of LUMO-I and LUMO-II) are consistent with the twofold-split "LUMO band" (~0.25 eV splitting) seen in UPS measurements of Cs-doped $C_{58}$ films.

***3.2.2 Cage-to-substrate coupling in submonolayer films.*** We were unable to obtain reproducible STS spectra of isolated monomers (adsorbed cages often jumped out of the imaged region at high bias tunnelling conditions). However, stable STS profiles could be obtained for surfaces covered by a nearly saturated monolayer. Fig. 4b shows LDOS($E_B$) curves measured at five points (chosen at random) over such an adsorbed $C_{58}$ submonolayer. All spectra show contributions from both the substrate and from $C_{58}$-derived states (which correlate with those seen for thick films) – in different relative amounts. Apparently, cage-surface interactions lead to a significant, site-specific narrowing of the HOMO-LUMO gap, which is observed to vary from 0.9 to 0.4 eV depending on orientation and lateral environment (spectra a-b). At some positions, no gap could be identified at all - the region probed appears to be conducting (spectra d-e). Similar phenomena have been observed for $C_{60}$ layers deposited onto Ag and Au surfaces /[27,32]/. In the case of $C_{60}$ on Ag films, the gap drops from 3.4 eV to 1.8 eV when going from a fullerene trilayer to a monolayer. For $C_{60}$ monolayers on Au(111) a by 1.2 eV reduced band gap has been observed.

Our observations for submonolayer $C_{58}$ films, can be qualitatively reproduced by DFT calculations for individual $C_{3v}$ and $C_s$ cages differently oriented on Au(111) (lower panel in Fig. 4b). The calculated LDOS functions exhibit pronounced narrowing of the HOMO-LUMO gap as compared to measurements for a thick $C_{58}$ film (for some orientations narrowing down to ~0.3 eV). Consequently, both theoretical and experimental $DOS(E_B)$ functions indicate a considerable hybridization of the HOMO and LUMO states as a result of interaction with the substrate (details of the adsorbate-substrate configurations calculated are presented in appendix).

***3.2.3 Mapping states in multilayered oligomers.*** Height profiles taken across the aggregate shown in Fig. 5 (mean height around 0.6 nm, see Fig. 5a), reveal that it is composed of at least two cage layers with a corresponding volume sufficient to enclose more than eight $C_{58}$ cages (with three interconnected cages on top). STS measurements at a constant $eU_{bias}$ value matched to the binding energy of a particular electronic state can be used to map the lateral distribution of that state /[33]/. This was done to generate the LUMO- and HOMO-maps of the $C_{58}$ aggregate (at $eU_{bias}$ of -0.51 and +0.64 eV, respectively) shown in Fig. 5c and 5d. Interestingly, while both HOMO and LUMO exhibit intense *stripe-like* features, their intensity distributions and relative orientations are qualitatively different (also compared to the corresponding topographic image).

Again, these orbital images do not provide enough information to reconstruct the cage structures. However, the experimental LUMO-map, in particular its on-top trimer segment (see Fig. 5d), strongly resembles the LUMO orbital distribution as calculated for a covalently interlinked $(C_{58})_3$ chain shown in Fig. 5b. Note that a similar (stripe-like patterned) distribution of HOMO and LUMO states has been calculated for covalently interlinked $[C_{60}]_3$ trimer /[34]/).

**4. Summary**

We have used 4.2 K STM to study deposits formed by exposing clean Au(111) surfaces to a low energy ion beam of mass selected $C_{58}^+$ at room temperature. $C_{58}$ molecules survive soft-landing ($E_{kin}$~6 eV) on Au(111) as intact non-IPR fullerene cages. Growth of $C_{58}$ deposits proceeds in three stages: (1) pinning by point dislocations at the elbows of the periodically ordered heringbone reconstruction (as well as pinning by step edges), (2) 2D island growth nucleated by the defect-immobilized $C_{58}$ cages (as mediated by covalent intercage bond formation) and (3) onset of 3D growth before wetting of the first layer is completed.

For thick $C_{58}$ films an effective HOMO-LUMO gap of 1.2 eV has been derived from STS-data. At submonolayer coverage, a considerable narrowing of this gap has been found which reveals a strong interaction with the underlying gold atoms. Covalently linked oligomers (both those in direct contact with the substrate and those offset from it), manifest characteristic STM topographic images with a stripe-like intensity pattern oriented perpendicular to the axis connecting the cage centres. In analogy to STM images of polymerized $C_{60}$ chains, these structural motifs can be considered as STM-specific markers of covalent inter-cage bonding. Our DFT calculations also support this assignment.

While statistical sampling was not performed, the limited number of topographic images obtained imply that LECBD of $C_{58}^+$ onto Au(111), leads preferentially to the covalent interlinking of deposited cages under our conditions. In contrast, STM-studies of 2D island films of metalloendofullerenes, $La_x@C_n$ (n<84), sublimed onto H-Si(100)-(2×1) /[35]/ show little evidence for stripe-like features indicative of interlinked monomers - although the material also comprises non-IPR cages. Similarly, covalently linked oligomers were also absent in an STM study of non-IPR $Dy@C_{68}$ deposited onto Ag ($\sqrt{3}$x$\sqrt{3}$)/Si(111) /[36]/. Apparently, nIPR motifs by themselves are not sufficient to ensure cage polymerization. Note however, that the corresponding $M@C_n$ cages are much larger, less stained and therefore less reactive than $C_{58}$ /[37]/. Furthermore, encapsulated metal atoms can induce rehybridization of nearby carbon atoms thus potentially "quenching" the outer reaction centre. In future it will be of interest to explore whether LECBD conditions can be found (e.g. by deposition into inert gas matrix followed by careful thawing) which also preclude covalent interlinking of non-IPR $C_{58}$.

## 5. Acknowledgement

The authors acknowledge support by the DFG Center for Functional Nanostructures (CFN).

**Figure Captions**

*Fig. 1 (color online)*
Typical STM topographic images after depositing $C_{58}$ at low coverage (~5 x $10^{12}$ cm$^{-2}$) onto Au(111).
(a) Overview scan [12.3 x 29.6 nm]: individual $C_{58}$ cages appear as bright circular spots preferentially occupying the point dislocation sites, Au(#). Line S marks a dislocation line which separates the fcc and hcp stacking regions. Line D indicates the domain walls. STM imaging parameters: $U_{bias}$=0.6 V, I=36.4 pA, $V_{scan}$=133 nm/s, τ=1.17 ms/point.
(b) Two representative topographic images taken for $C_{58}$ cages pinned at point dislocation sites, $C_{58}$-Au(#) (left panel) in comparison to an analogously pinned $C_{60}$ cage (right panel). The corresponding height profiles are shown underneath.
(c) The upper row shows topographic images for individual $C_{58}$ cages adsorbed on flat terrace regions (i.e. not pinned at Au(#) sites). In the lower row DFT-calculated iso-density surfaces for differently oriented $C_{58}(C_s)$ and $C_{58}(C_{3v})$ monomers are shown for comparison to the experimental images. The configurations and the basis sets are indicated (for calculation details see appendix). Note the generally good agreement between STM topographic images and DFT-based simulations (vertically aligned image pairs from (a) to (e))

*Fig. 2 (color online)*

Typical STM topographic images observed at intermediate $C_{58}$ coverages (~5 x $10^{13}$ $C_{58}$/cm$^{-2}$):

(a) Overview scan showing small 2D islands predominantly pinned by Au(#) sites [area: 80 x 80 nm].

(b) Topographic images of two $C_{58}$ dimers (left panel) and a linear $C_{58}$ trimer (right panel). The corresponding line profiles are shown in the lower row. Imaging parameters were analogous to those given in the figure 1 caption.

(c) Model $C_{58}$ dimer geometry as used for DFT simulations of constant-current images (left panel) consisting of two covalently interlinked cages with one (pinned) cage attached to the surface via its heptagonal ring. DFT calculations of the corresponding density iso-surfaces (two images on the right) were performed for two tip-cage distances, ~3 and ~4.5 A (corresponding to local densities of states of $2 \times 10^{-6}$ and $10^{-19}$ au$^{-3}$, respectively). The predicted topographic image is dominated by horizontal bright stripes oriented perpendicular to the axis connecting the cage centers. This finding qualitatively reproduces typical STM topographic images assigned to dimers (middle panel).

*Fig. 3 (color online)*

Typical STM images illustrating the surface topography after depositing a high $C_{58}$ dose of ~2 x $10^{14}$ cm$^{-2}$ onto Au(111) (Imaging parameters: I= 0.59 nA, $U_{bias}$= 301 mV (left images), I= 0.63 nA, $U_{bias}$= 305 mV (right)). Film growth proceeds via formation of three dimensional islands consisting of predominantly $C_{58}$ oligomers. The lateral density of these pyramidal islands corresponds roughly to that of the underlying point dislocation sites (Au(#)). Height profiles (e.g. underneath left panel) indicate that their heights is ~4 ML. Note the well distinguishable *stripe-like* motifs indicating covalent interlinking of the cages to form oligomeric $C_{58}$ chains.

*Fig. 4 (color online)*

(a) Local densities of states LDOS~dI/dV/(I/V), as obtained from STS measurements at seven different sites over the outermost layer of a thick $C_{58}$ film. The DOS profiles change when going from site to site but the main peaks remain nearly at the same positions: LUMO-I ($E_B$= 0.35 eV), LUMO-II ($E_B$= 0.5 eV), HOMO-I ($E_B$= -0.85 eV) and HOMO-II ($E_B$=-1.3 eV). The mean distance between the HOMO-I and LUMO-I peaks defines the band gap of 1.2 eV. The corresponding measurement conditions were: $U_{bias}$=309 mV and I=0.55 nA, scan range: -1.3-0.7V, 90 spectra within the bias range (-1.3V ÷0.7 V), τ= 250 μs/point.

(b) Local densities of states, LDOS($E_B$) obtained by STS at five sites over a $C_{58}$ submonolayer (upper panel). The DOS profiles change considerably when going from site to site. New states appear within the 1.2 eV-wide gap found for thick films. The LDOS($E_B$) calculated for four different orientations of $C_{58}$ isomers ((1) $C_{58}(C_s)$He, (2) $C_{58}(C_{3v})$H, (3) $C_{58}(C_s)$Pe, (4) $C_{58}(C_{3v})$P) reproduce the main trends (see text and appendix for details).

*Fig. 5 (color online)*

(a) STM topographic image and corresponding height profile of a $C_{58}$ aggregate consisting of at least eight cages: five cages in the lower layer and three cages "on-top" (imaging parameters comparable to those given in figure 1).

(b) For comparison we also show here the spatial distribution of LUMO-states as calculated for an oligomer consisting of three covalently interlinked $C_{58}$ cages. Its structure has been optimized using the SVP-basis set. The colors distinguish the relative sign of the wave function of a LUMO-orbital (for details see appendix).

(c) Spatial distribution of HOMO-states for the $C_{58}$ aggregate (imaging parameters: area 8 x 8 nm, $\delta z$= 73.2- 477 mV, offset from -19.0 to 52.3 nm, $U_{bias}$=-0.51 V, I=1.86 nA, $V_{scan}$= 8 nm/s, $\tau$=3.91 ms/point).

d) Corresponding map of the LUMO-states (imaging parameters: area 8 x 8 nm $\delta z$= 15.6-301 mV, offset from -19.0 to 52.3 nm, $U_{bias}$=0.65 V, I=1.97 nA, $V_{scan}$=8.00 nm/s , t=3.91 ms/point)

**Appendix**

DFT-calculations of the geometry and the electronic structure have been performed using the quantum chemistry package TURBOMOLE [18]. A generalized gradient approximation for the exchange-functional (GGA, BP86 functional [38,39]) has been employed. Dispersion effects have been included using DFT-D3-type Grimme corrections [40]. All calculations have been performed with at least two different sizes of contracted Gaussian-type basis sets: the split-valence type (def-SVP) and triple-zeta type (def-TZVP) basis sets including polarization functions [41,42]. Corresponding Coulomb-fitting basis sets have been used within the resolution of identity (RI) approximation [43]. To investigate convergence with respect to the basis set size, we only varied the basis set for the carbon atoms but kept the basis set for gold fixed to def-SVP. First the adsorption geometries of $C_{58}$-monomers were explored [25]. For two lowest energy configurations in vacuum, a $C_s$- and a $C_{3v}$-symmetric version, 9 and 18 different adsorption geometries have been investigated, respectively (see Fig. A1). The adsorption geometry not only depends on the polygon facing the surface but also on the lateral positions of the carbon atoms with respect to the positions of the gold atoms. After a pre-relaxation, the monomers have been structurally optimized in every adsorption geometry separately. Therefore, the monomers are placed on a (fixed) Au(111)-surface which is modeled by two (finite) layers of gold atoms using absorbing boundary conditions to include the effects of the rest of the gold-substrate [44]. For all configurations, the binding energy of the monomer to the surface has been calculated for several binding distances to the Au(111)-substrate.

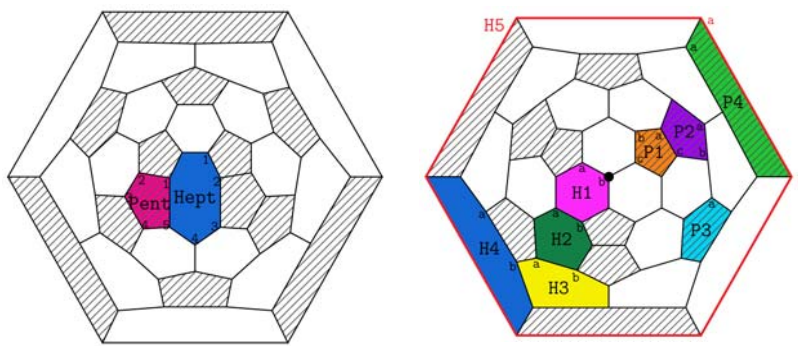

**Figure A1**. Schlegel diagrams of two $C_{58}$-monomers showing different adsorption geometries. For each configuration, the corresponding colored polygon was facing the surface with the labeled carbon atom sitting in an on-site position. All pentagons are shaded to facilitate finding reactive centers violating the IPR-rule. Left: $C_s$-monomer showing 9 different adsorption geometries. Since this monomer contains three adjoining pentagons (violating the IPR-rule), the most reactive bonds are assumed to be either on the heptagon or on the colored pentagon. Other polygons facing the gold surface will likely have a weaker binding and have not been investigated. Right: $C_{3v}$-monomer showing 18 different adsorption geometries. The three-fold rotation axis is denoted by the black dot in the center.

As strongest binding to the gold surface, we found the $C_s$-monomer with a binding energy of 4.16 eV when the heptagon is facing the surface in such a way that the carbon-4 atom is in a distance of 2.27Å showed good convergence with respect to the basis set size. Choosing other heptagon atoms (1-3) to be on-site slightly varies the binding energy (down to 3.95 eV) but the binding is still stronger than compared to the pentagon facing the surface. The binding of the $C_{3v}$-monomer is slightly weaker. The binding energy is maximal for the hexagon H1 facing the surface. The on-site atom hardly plays any role; the binding energy only varies not significantly (4.03 eV for a-atom and 4.01 eV for b-atom). Because of the large symmetry group, having any of the other four carbon atoms on-site is equivalent to having a- or b-atom on-site.

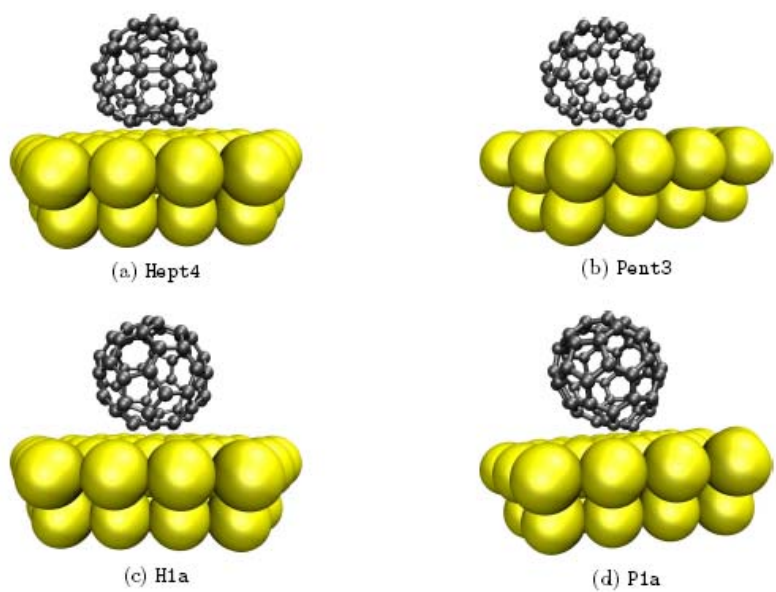

(a) Hept4  (b) Pent3  (c) H1a  (d) P1a

**Figure A2.** Four examples of the optimized adsorbate-substrate configurations. The labeling indicates the carbon ring contacting the substrate (as indicated in Fig. A1)

In second step, we investigated the adsorption of dimers formed by $C_{3v}$ and $C_s$ monomers [45]. Since there are many different possibilities of forming a dimer and adsorbing it on Au(111), we restricted ourselves to six configurations that are chosen by intuition. The two most promising configurations with respect to the experimental STM-images are the dimers which are formed using the following reaction mechanism: First, a monomer adsorbs in its favored orientation as calculated in the binding energy study above. Then, a free monomer docks to the already bound cage. In the $C_s$-case, this means that the first fullerene adsorbs with the heptagon to the gold surface whereas the second fullerene docks with its heptagon to the first fullerene. In the $C_{3v}$-case, the second fullerene can also adsorb with the H1-hexagon because there is a reactive bond located between two nIPR-pentagons which can be used for forming the dimer (see Fig. A3). The other four dimers follow a compromise, maximizing the adsorption energy vs. maximizing the binding energy. For further details about the other four configurations and for information of the transport theory used to calculate the STM-images see Ref. [17].

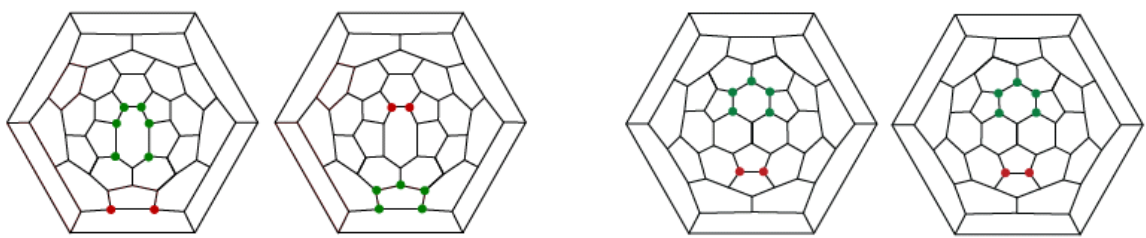

**Figure A3**. Schlegel graphs of the binding geometry of the two considered dimers. The red markers indicate atoms forming the bond between the two $C_{58}$-fullerenes. The green markers denote the carbon atoms in the immediate proximity to the gold surface. Left: a dimer consisting of two $C_s$-monomers. The symmetry group of the combined system is still $C_s$. Right: a dimer consisting of two $C_{3v}$-monomers. Because of binding and adsorption, the combined system exhibits $C_{2v}$-symmetry [45].

For all six configurations, the dimers have been structurally optimized in vacuum and then placed on the Au(111)-surface which again is modeled by two (finite) layers of gold atoms including absorbing boundary conditions [17]. For the exact placement, we used the adsorption orientation and distance calculated for the monomers. Our tests showed that further optimization of the structure placed on the gold surface does not lead to visible changes in the STM-image. Again, we also ensured that the features in the simulated STM-images do not depend on the basis set. They are virtually identical for SVP- and TZVP-basis set.

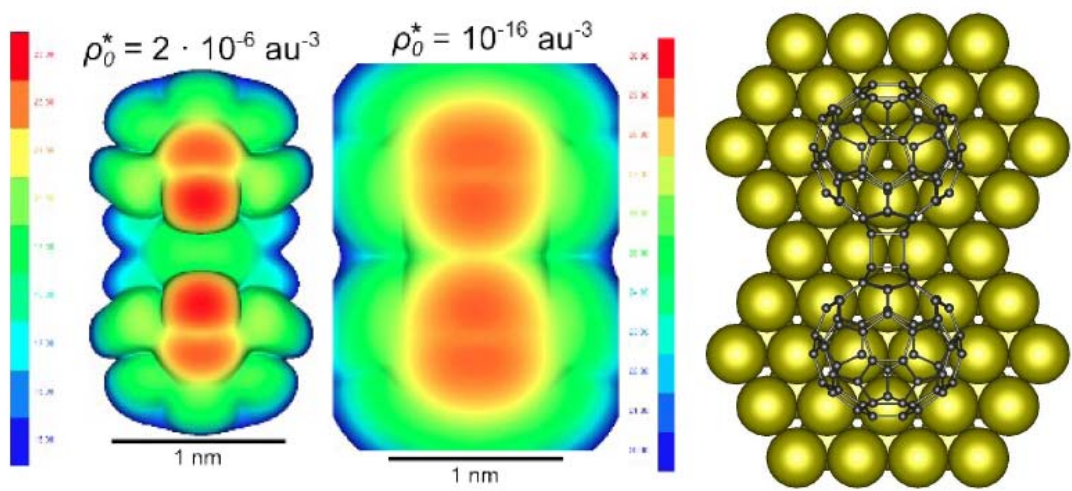

**Figure A4**. Left panel shows two iso-density surfaces obtained for a dimer consisting of two $C_{58}(C_{3v})$ cages interlinked via double C-C intercage bonds associated with the 2AP-2AP motif. The related adsorption geometry is shown in right panel. Again the topographic iso-surface is dominated by horizontal bright stripes oriented perpendicular to the axis connecting the cage centers.

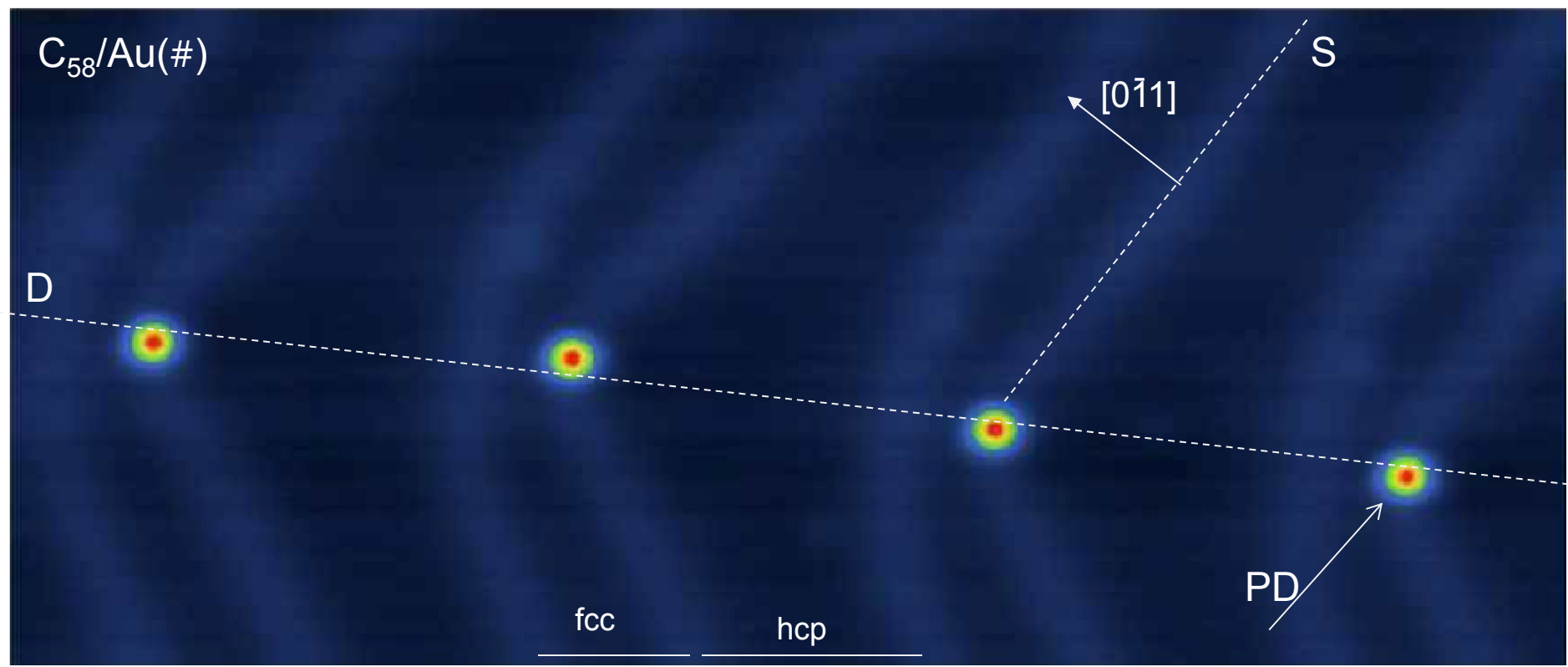

Fig. 1a  N. Bajales et al.

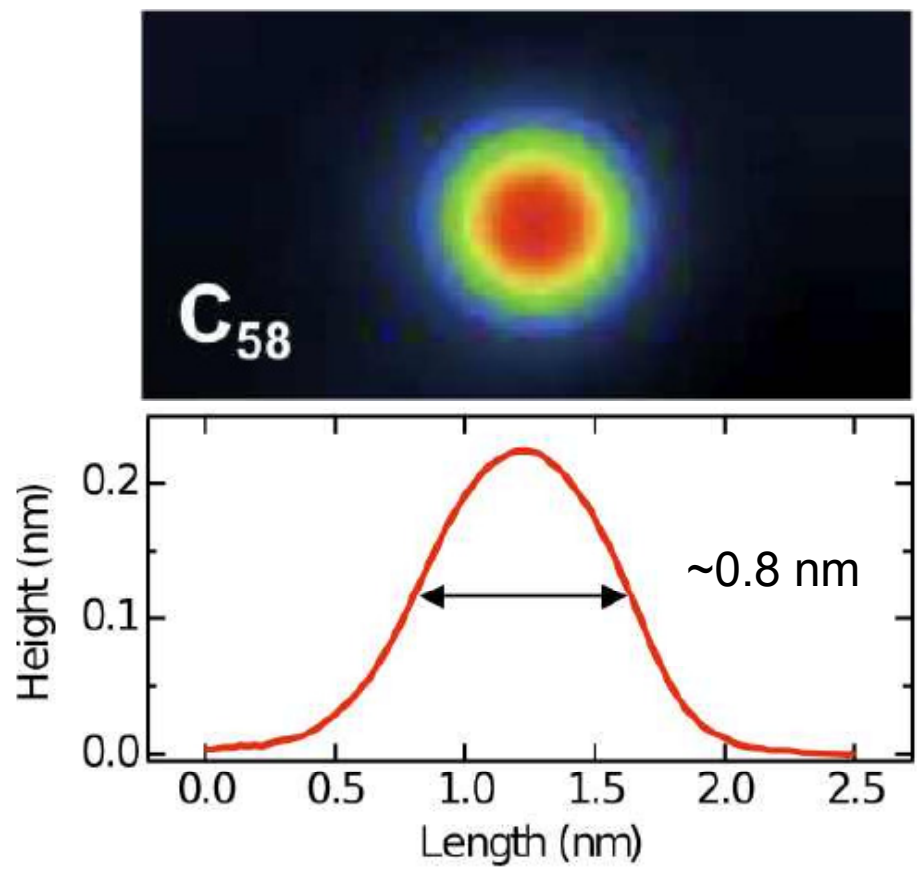 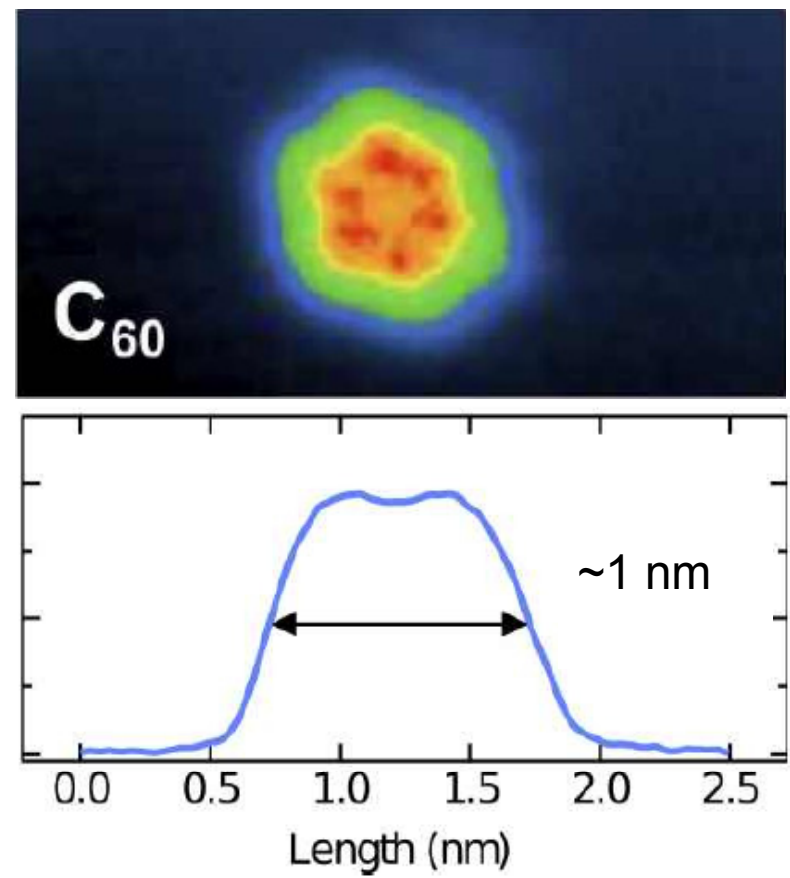

Fig. 1b   N. Bajales et al.

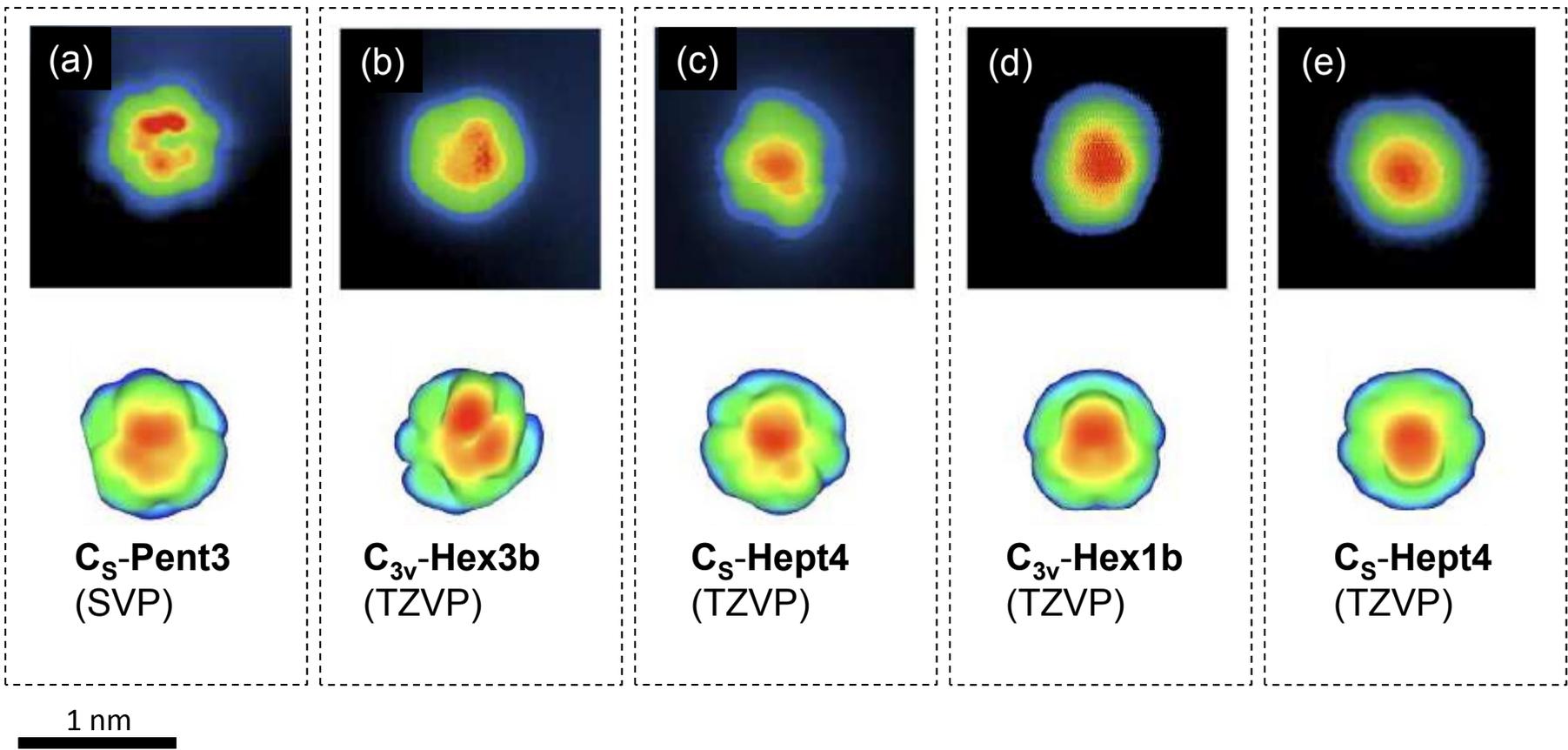

Fig. 1c  N. Bajales et al.

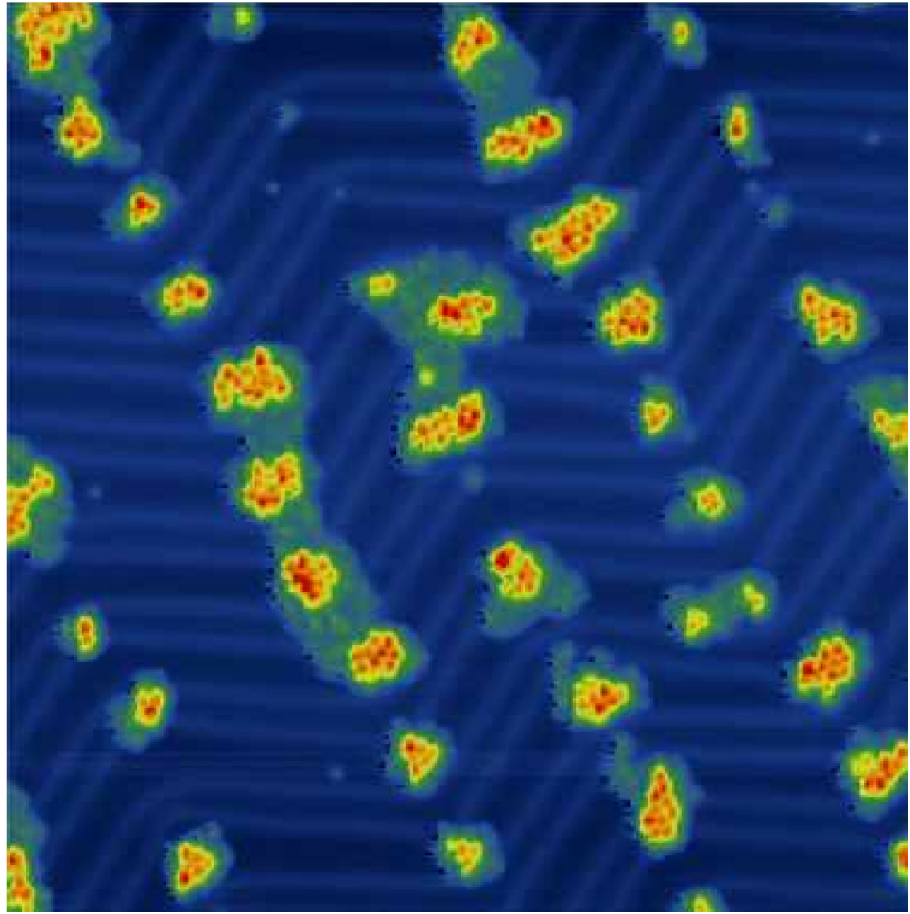

Fig. 2a   N. Bajales et al.

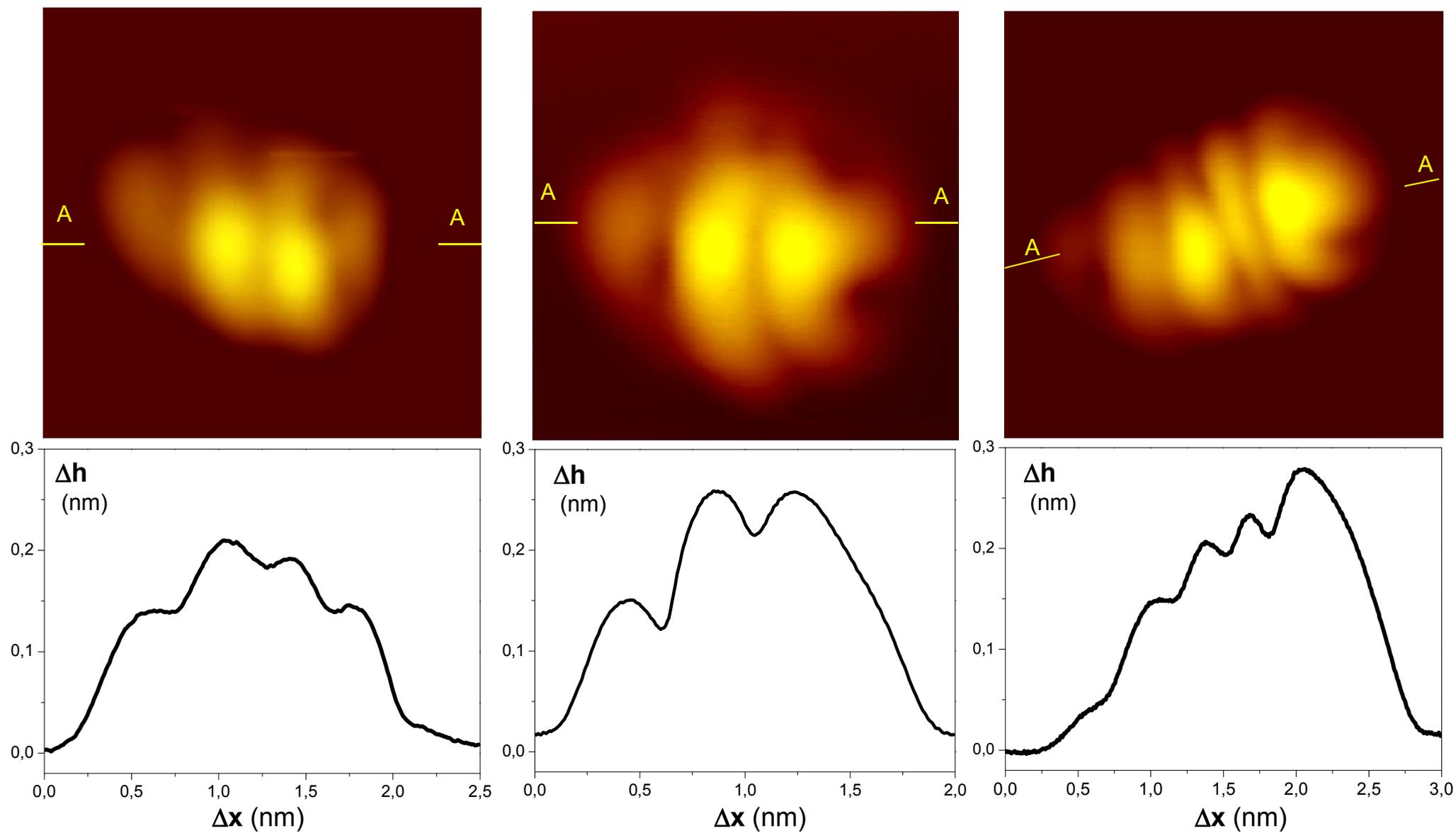

Fig. 2b   N. Bajales et al.

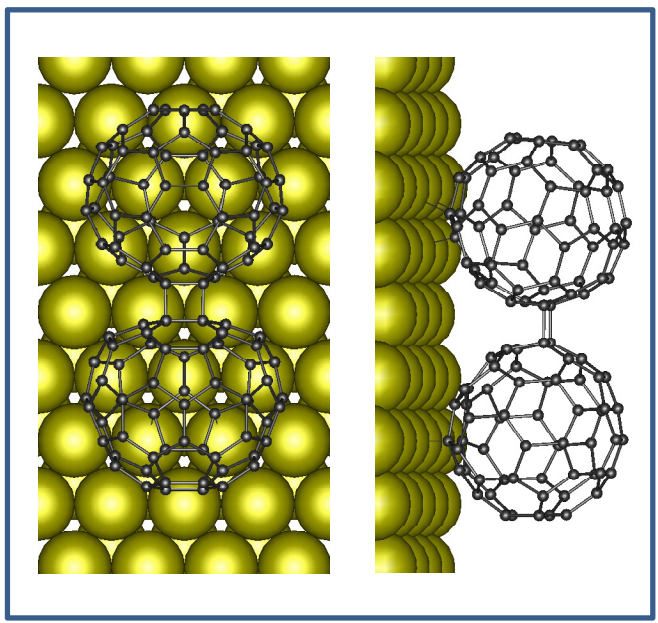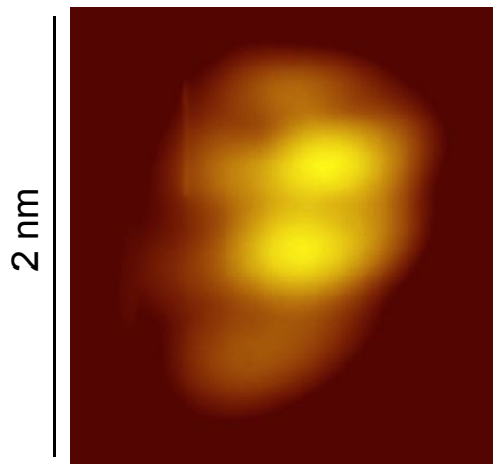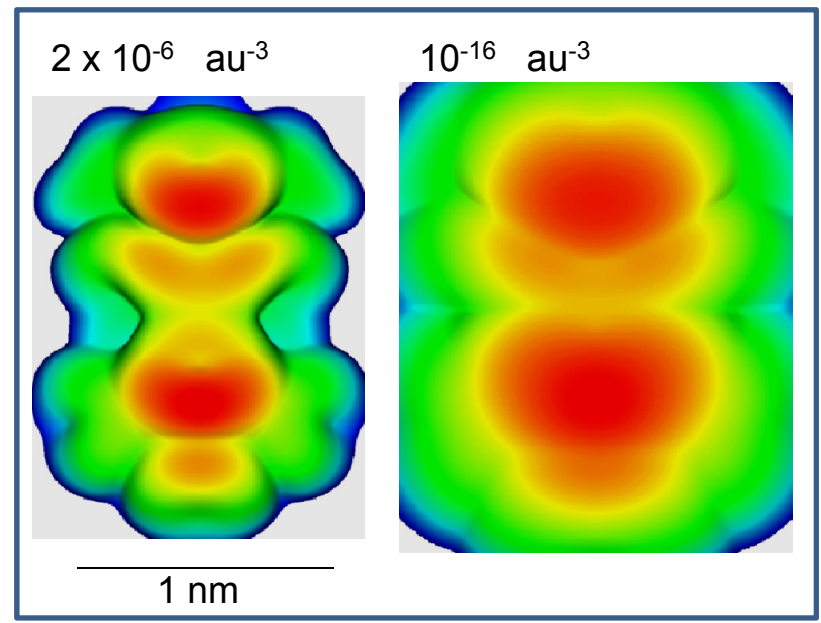

2 nm

2 x 10⁻⁶  au⁻³         10⁻¹⁶  au⁻³

1 nm

Fig. 2c   N. Bajales et al.

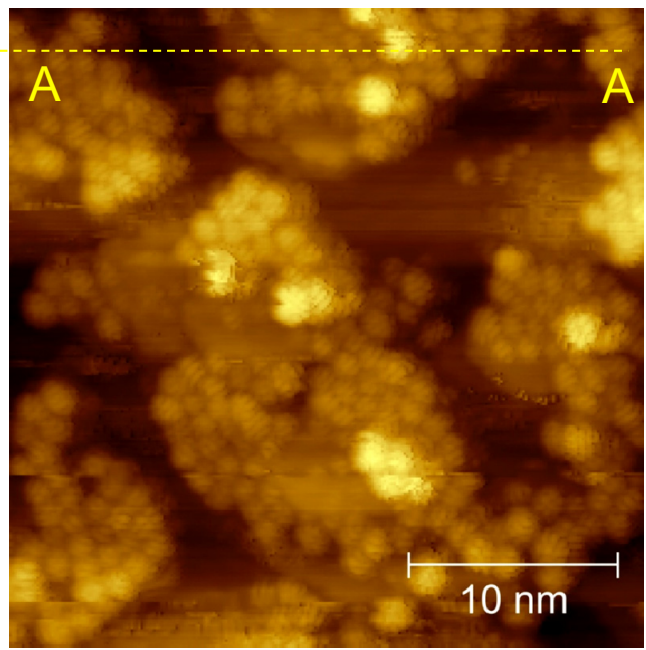 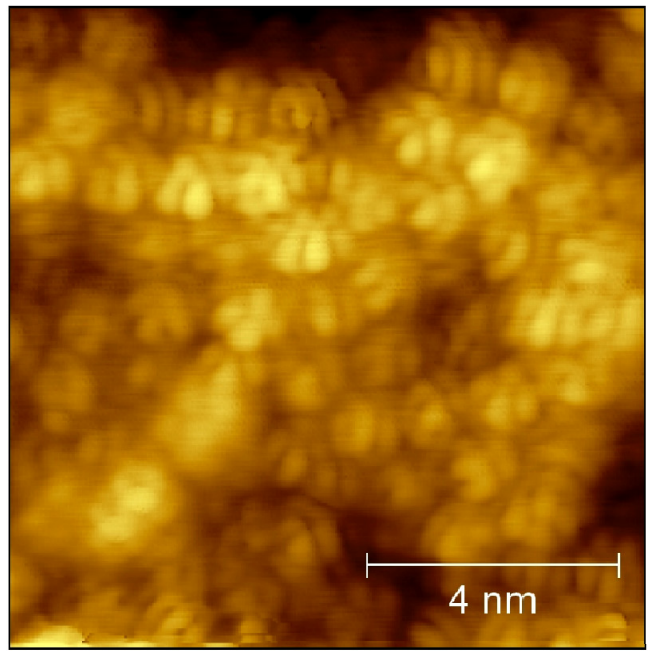 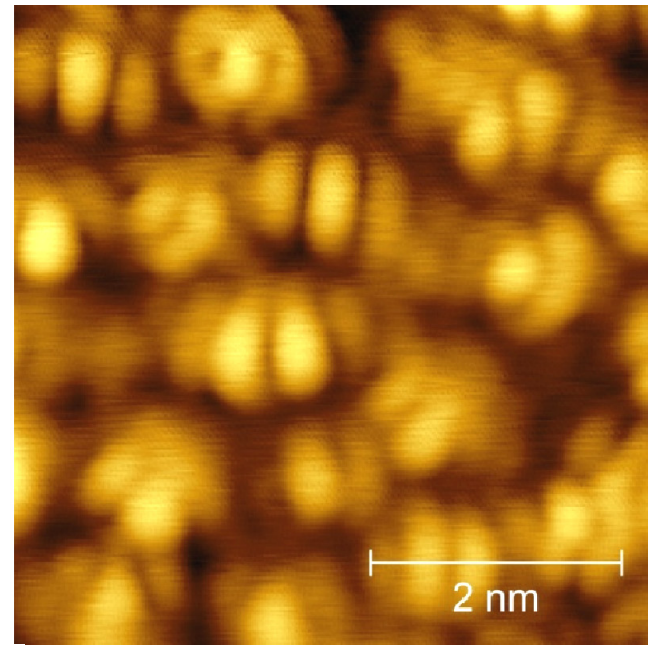

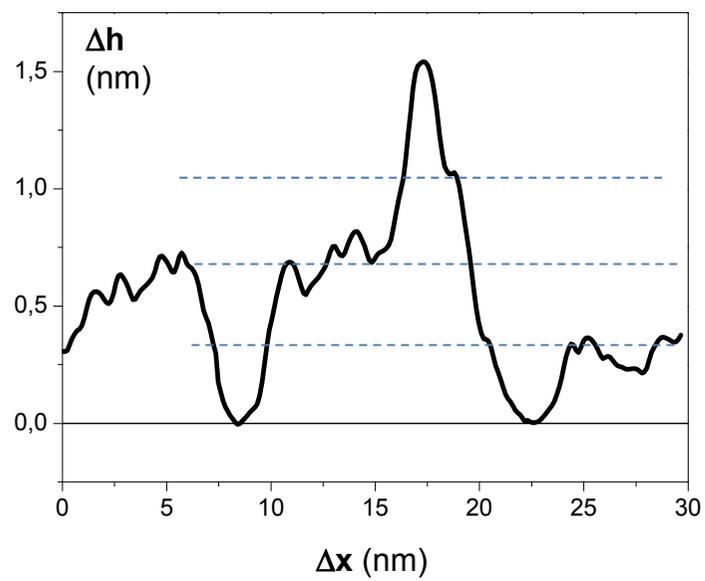

Fig. 3  N. Bajales et al.

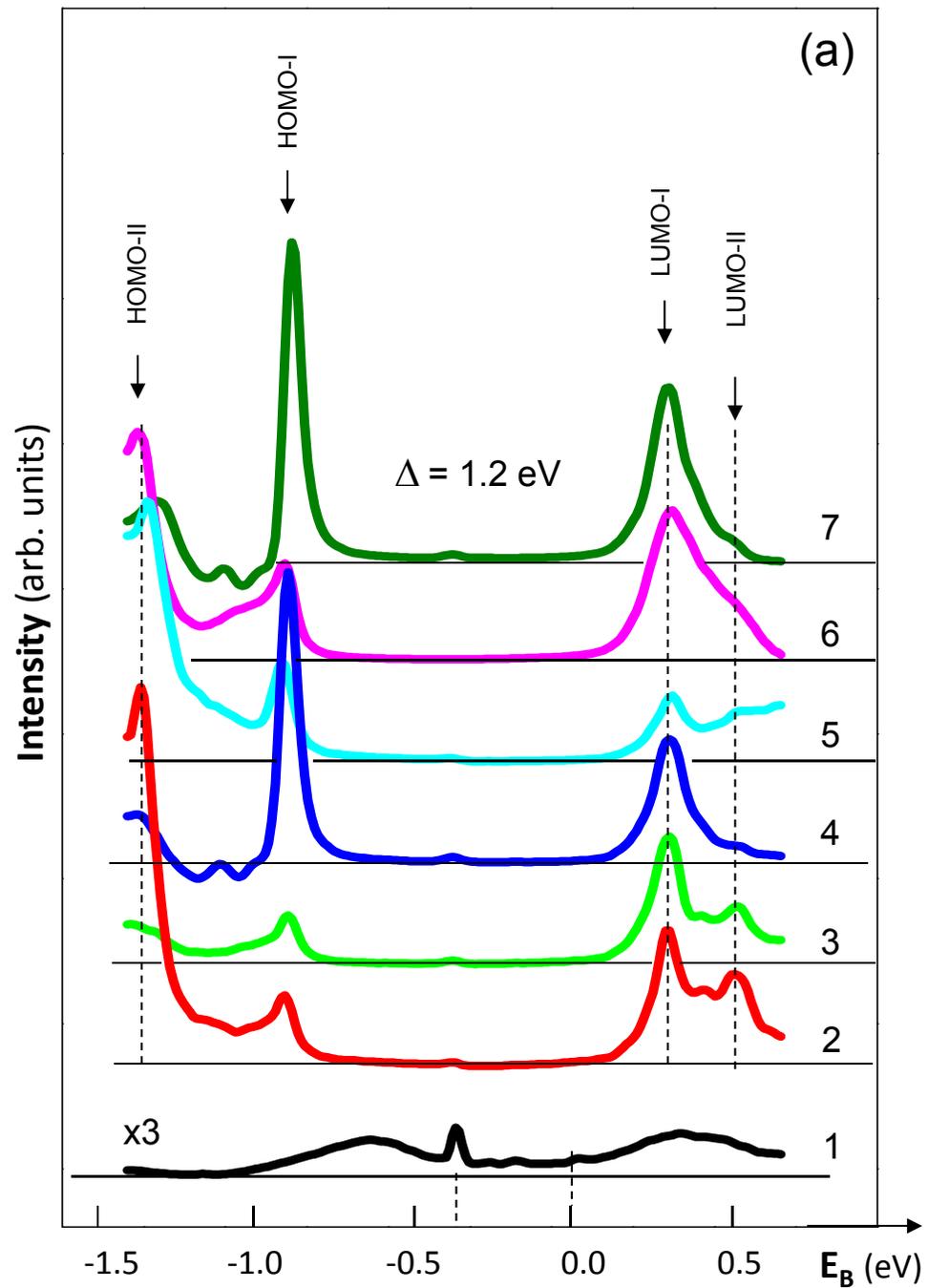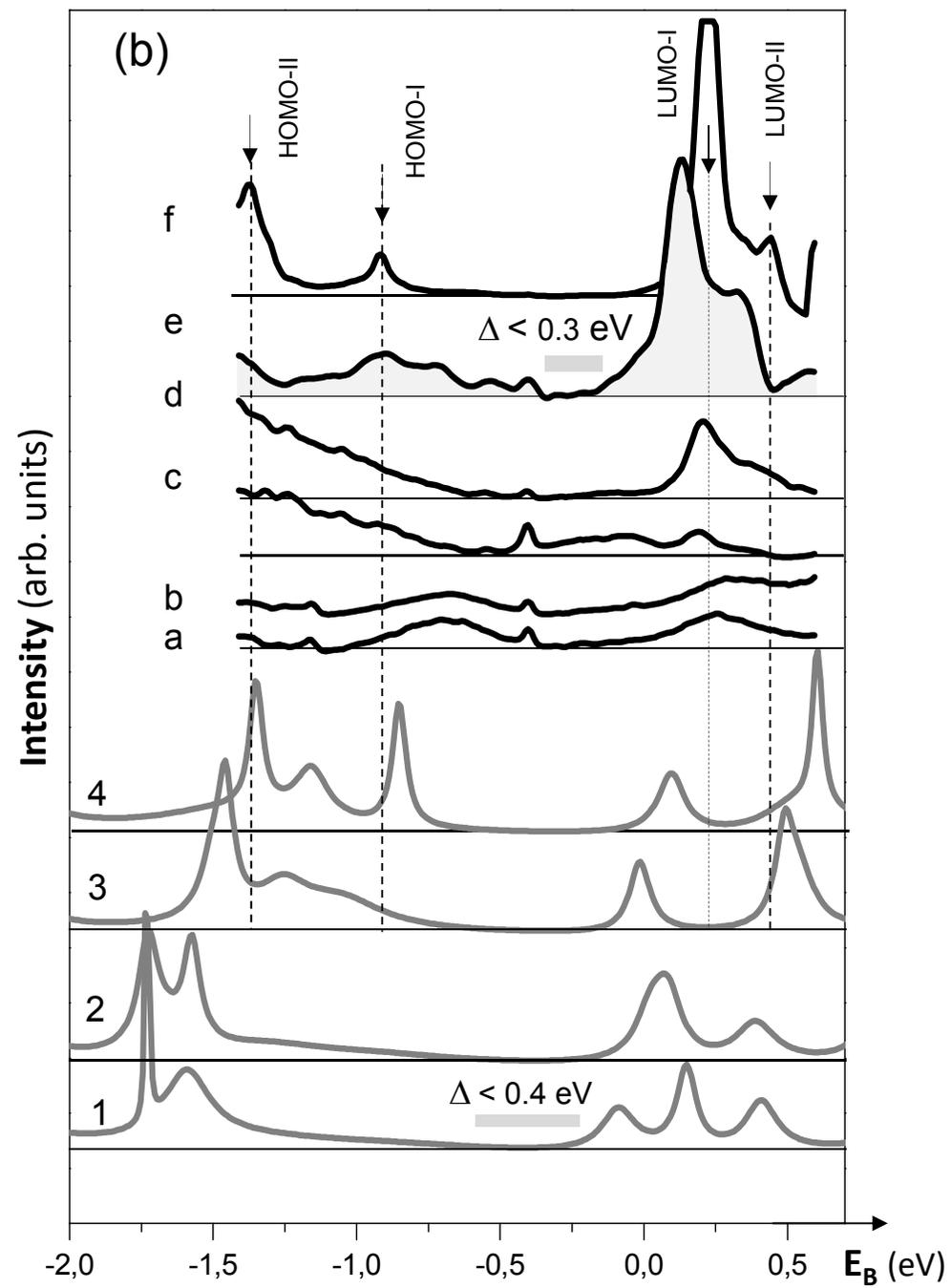

Fig. 4  N. Bajales et al.

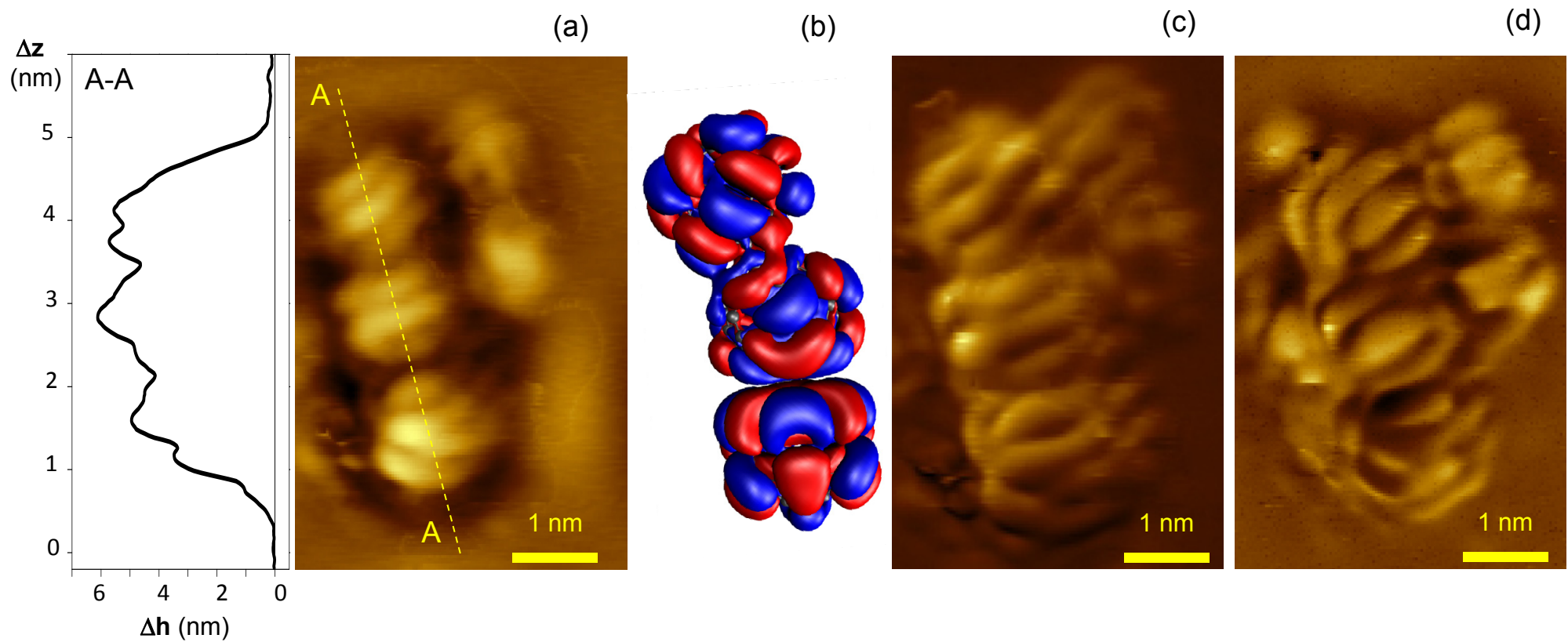

Fig. 5  N. Bajales et al.